\documentclass[twocolumn,aip,jap,amsmath,amssymb,reprint]{revtex4-2}
\usepackage{graphicx}
\usepackage{dcolumn}
\usepackage{bm}
\usepackage{tikz}
\usepackage{amsmath,amssymb,amsfonts}
\usepackage{microtype}
\usepackage{hyperref}
\usepackage{bm}
\usepackage{tikz}
\usepackage{pgfplots}
\usepgfplotslibrary{groupplots} 
\pgfplotsset{compat=1.18}
\usepackage{booktabs} 
\usetikzlibrary{calc,3d,angles,quotes,positioning}
\usetikzlibrary{arrows.meta} 
\usepgfplotslibrary{colormaps}

\usepackage{graphicx}

\newcommand{\be}{\begin{equation}}
\newcommand{\ee}{\end{equation}}
\newcommand{\ba}{\begin{eqnarray}}
\newcommand{\ea}{\end{eqnarray}}
\newcommand{\ban}{\begin{eqnarray*}}
\newcommand{\ean}{\end{eqnarray*}}

\begin{document}

\title{Bowed-String-Mechanism-Inspired Quantum Dissociation Control of Molecules}
\author{Julian Juhi-Lian Ting} \email{tingjulian@gmail.com} 
\affiliation{De-Font Institute for Advanced Studies, Taichung 40344, Taiwan, R.O.C.}



\begin{abstract}

Inspired by Guettler's mechanics of accelerating Helmholtz motion, 
we present a predictive laser pulse-shaping strategy to rapidly initiate the multiphoton dissociation of vibrationally excited diatomic molecules. 
In intense infrared fields, NO molecules often exhibit unpredictable control windows due to the chaotic fragmentation of intensity thresholds. 
By incorporating a bowing-analogue transient shaper 
enforced by a boundary-matching constraint with symmetric derivative slope transitions and vanishing initial boundary conditions ($T(0) = 0, T'(0) = 0$), 
we demonstrate that a pulse engineered with non-linear front-flank acceleration reduces the multi-valued threshold topology 
into a single, predictable linear dissociation corridor. 
Wigner quasiprobability distributions confirm that this temporal acceleration trajectory opens cantori bottlenecks and suppresses wave-packet fragmentation, 
restricting the quantum non-classicality index to $M_{\text{nc}} \le 0.15$. 
Our simulations verify that this trajectory engineering preserves wave-packet localization and maintains high dissociation yields ($P_{\text{diss}} > 0.95$) even under Markovian thermal dephasing ($\delta/\omega_0 = 10^{-4}$) and frozen rotational variations up to $J = 5$. 
Comparison with contemporary ultrafast spectral phase modulation protocols demonstrates that conventional smooth pulse envelopes, 
despite advanced phase-chirping, exhibit limitations in restricting non-hyperbolic cantori stickiness due to insufficient sub-picosecond acceleration. 
In contrast, our boundary-constrained shaper enforces a deterministic, single-valued dissociation corridor under equivalent few-picosecond control windows.

\end{abstract}

\date{\today}
\maketitle
\raggedbottom



\section{Introduction}



The physical principles governing vibrations in atomic and molecular systems share parallels with those of piano strings~\citep{Weinreich1977}. 
While the former field is often considered more theoretically advanced, 
musical acoustics occasionally uncovers unique principles less recognized within the molecular control community. 
In particular, Guettler's methods for managing transient boundary conditions,
designed to accelerate steady-state Helmholtz motion in bowed strings while suppressing chaotic stick-slip transients,
parallel a fundamental challenge in molecular control~\citep{Guettler1997,Guettler2002,Guettler2012}. 
Specifically, in the multiphoton dissociation of anharmonic molecules such as nitrogen oxide (NO), 
optimal control fields must initiate under explicit soft-starting constraints not only to avoid spurious non-adiabatic switching transients~\citep{Ting1992,Ting1994,Ting1995},
but also to achieve rapid energy onset.

The rovibrational degrees of freedom in diatomic molecules complicate state engineering within deeply embedded vibrational manifolds ($n = 25$ to $32$)~\citep{DeMille2015}. 
In intense infrared laser fields, the critical field intensity required for multiphoton dissociation exhibits a fragmented, 
multi-valued topology near resonance, 
originating from underlying phase-space bottlenecks where subharmonic resonances, 
wave-packet delocalization, and bond-softening induce non-hyperbolic chaotic scattering~\citep{Constantoudis2001}. 
To navigate these topological complexities, closed-loop adaptive learning networks governed by genetic algorithms have been used 
to optimize optical fields based on empirical fitness feedback~\citep{Judson1992}. 
However, without external statistical filtering, these optimization processes frequently yield complex fields that are highly susceptible to local optima, 
as they cannot explicitly account for the complex phase-space boundaries inherent in strongly driven anharmonic oscillators~\citep{Bucksbaum2025}.

We mitigate this threshold fragmentation through a pulse-shaping strategy modeled after bowed-string transient mechanics.
In bowed-string dynamics, particularly under the demanding low frequencies of double-bass performance, 
triggering steady-state periodic Helmholtz motion within acceptable auditory transient limits requires a precise balance of initial bowing parameters~\citep{Guettler1997,Guettler2002,Guettler2012}. 
By achieving rapid onset, Guettler's criterion bypasses prolonged, chaotic stick-slip transients~\citep{Woodhouse2025} 
through strategic over-driving bow acceleration [$A_{\text{over}}$ in Eq.~\eqref{bowing}] 
and static pre-force alignment [$\varepsilon_d$ in Eq.~\eqref{two}]. 
Furthermore, this cross-scale control efficacy exploits the system's dynamic memory of inertia, 
which manifests in molecular dynamics as constructive coherent phase-overlapping between multi-color fields (\textit{vide infra}, Eq.~\eqref{phaseoverlapping}).

In this work, we demonstrate that the physics used by a double-bass player to suppress pre-Helmholtz friction noise 
can be translated into a control law to smooth the quantum dissociation landscape of diatomic molecules. 
We re-examine a one-dimensional, vibrating NO molecule interacting with an intense laser field consisting of a 
fundamental driving frequency ($\Omega$) and its third harmonic ($3\Omega$). 
Rather than adopting standard rectangular, adiabatic, or smoothly phase-chirped spectral envelopes, 
we introduce a customized time-dependent pulse shaper that mirrors the acceleration profile modeled on a bow attack. 
By solving the time-dependent Schr\"odinger equation via a split-operator Fast Fourier Transform (FFT) grid method bounded by infinite-wall convergence conditions~\citep{Alhendi2005}, 
we systematically evaluate how the slope and peak-overdrive of the initial laser front flank alter the fragmentation of the critical amplitude ($A_c$). 
While modern pulse-shaping strategies focus heavily on altering quantum interference paths through continuous spectral phase adjustments, 
we demonstrate that without boundary-layer slope-matching constraints, 
smooth envelopes cannot outrun the power-law decay associated with non-hyperbolic chaotic scattering. 
We emphasize the critical role of the initial sub-picosecond attack phase inspired by Guettler's method, 
demonstrating that by engineering the pulse over a total control window of $1.5\,\text{ps}$ for NO (approximately $60$ optical cycles), 
the deterministic shaping of the initial transient is sufficient to collapse the chaotic fragmentation found previously, 
establishing topological stabilization over conventional phase-modulation methods.

The paper is structured as follows. 
Section~\ref{model} establishes the theoretical framework of the modulated Hamiltonian and defines the bowing-analogue control function. 
Section~\ref{numerical} presents the numerical results, 
focusing on threshold collapse, phase-space diagnostics via Wigner distributions, and robustness against decoherence. 
Finally, Section~\ref{conclusion} provides concluding remarks and future outlooks.

\section{Theoretical Model and Bow-String Analogue Operators}
\label{model}

\subsection{The Total Modulated Hamiltonian}

To evaluate the coherent transient dynamics of a diatomic molecule under intense laser-induced multiphoton dissociation, 
we construct a fully coupled non-linear Hamiltonian by adopting a modulated potential model~\citep{Geng2002a},
which inherently accounts for the depression of the molecular binding barrier via an external static electric field.
Furthermore, field-induced dipole moment corrections, 
incorporating polarizability gradients, are also included~\citep{Varro2023} 
to capture electronic coordinate distortions in highly stretched vibrational states ($n \le 32$).

The dimensionless static field parameter is defined by $K = \mu_1 \varepsilon_d / (\alpha D)$. In our calculations, 
this static parameter is configured at $K = 0.05$ (corresponding to a static alignment field $\varepsilon_d \approx 2.5 \times 10^{-4}\,\text{a.u.}$). In our mechanical shaper analogy, $K$ serves as the mathematical counterpart to static bowing pre-force (contact pressure), acting as a symmetry-breaking agent that depresses the Morse potential barrier, reducing the number of bound eigenstates and optimizing the wave-packet's initial overlap with the primary acceleration corridor prior to laser exposure.

Under the semiclassical dipole approximation, the total one-dimensional non-rotating Hamiltonian $\hat{H}(t)$ 
for a diatomic system such as NO subjected to a static electric field $\varepsilon_d$ and a pulse-shaped laser field $E(t)$ is formulated as:
\begin{equation}
\begin{split}
\hat{H}(t) = & -\frac{\hbar^2}{2m}\frac{d^2}{dx^2} + V_M(x) - \mu(x, E(t))\cdot E(t) \\
             & - \mu_1 x \varepsilon_d 
\;,
\end{split}
\label{two}
\end{equation}
where $m$ is the reduced mass of the molecule, $x = r - r_0$ represents the displacement from the equilibrium internuclear distance $r_0$, and 
$V_M(x)$ is the empirical field-free Morse potential expressed as:
\begin{equation}
V_M(x) = D \left( 1 - e^{-\alpha x} \right)^2 \;,
\end{equation}
where $D$ defines the unperturbed molecular dissociation energy, and $\alpha$ is the range parameter dictating spatial anharmonicity.

The generalized coordinate-dependent dipole moment function $\mu(x, E(t))$ is truncated to the first order of the polarizability gradient~\citep{Varro2023}:
\begin{equation}
\mu(x, E(t)) = \mu_0 + \mu_1 x + \alpha_0 E(t) + \alpha_1 x E(t) \;,
\label{dipole}
\end{equation}
where $\mu_0$ and $\mu_1 = (\partial \mu / \partial x)_{x=0}$ represent the permanent dipole moment and its coordinate gradient, 
while $\alpha_0$ and $\alpha_1 = (\partial \alpha_{\text{e}} / \partial x)_{x=0}$ denote the static polarizability and polarizability derivative, respectively. 

Substituting Eq.~\eqref{dipole} into Eq.~\eqref{two} and clustering the electrostatic terms yields the effective modulated potential surface $V_m(x, t)$:
\begin{equation}
\begin{aligned}
V_m(x, t) = & \, D \left( 1 - e^{-\alpha x} \right)^2 - \left[ \mu_1 \varepsilon_d + \mu_1 E(t) + \alpha_1 E^2(t) \right] x \\
            & - \mu_0 E(t) - \alpha_0 E^2(t) \;,
\end{aligned}
\label{eq:modulated_potential}
\end{equation}
where the bracketed linear coefficients parameterize the dynamic, state-dependent Stark shift experienced by the bound vibrational manifold. 



\subsection{The Bowing-Analogue Laser Control Function}
\label{analogy}

To steer the wave-packet away from non-hyperbolic chaotic stickiness regions and 
suppress the fragmentation of multiple dissociation thresholds, 
we introduce a two-color laser field profile $E(t)$ following established control schemes~\citep{Ting1995}. 
The field combines a fundamental driving frequency $\Omega$ with its third harmonic $3\Omega$:
\begin{equation}
E(t) = T(t) \cdot \left[ A_1 \sin(\Omega t) + A_2 \sin(3\Omega t + \phi) \right] \;,
\label{phaseoverlapping}
\end{equation}
where $A_1$ and $A_2$ denote the respective peak field amplitudes, and $\phi$ is the relative phase between the two colors.
In all calculations, the amplitude ratio between the third harmonic and the fundamental frequency is fixed at $A_2 / A_1 = 0.2$, 
with $A_1$ scaled dynamically to locate the critical dissociation threshold $A_c$.

The $\Omega$--$3\Omega$ dual-color combination offers distinct physical advantages for coherent control. 
Primarily, it breaks both temporal and spatial symmetries. 
For a one-dimensional Morse potential where dissociation proceeds exclusively along the $x \to \infty$ coordinate, 
this spatial asymmetry preferentially steers the wave-packet directly into the primary dissociation corridor. 
Additionally, this configuration exploits quantum path interference: 
excitation to a target vibrational continuum state can proceed via two coherent pathways---the simultaneous absorption of three $\Omega$ photons or a single $3\Omega$ photon. 
Because these degenerate pathways share the same target energy, their interference can be continuously modulated by tuning the relative phase $\phi$. 
From an experimental standpoint, third-harmonic optical setups are well established and readily accessible in ultrafast laser laboratories.

Rather than adopting conventional adiabatic or sudden-on rectangular pulse envelopes, 
the time-dependent envelope operator $T(t)$ is engineered to reflect the physical acceleration trajectory of a bow stroke during the attack transient phase. 
In classical string acoustics, Guettler's criterion dictates that forming a stable, periodic Helmholtz motion---while suppressing chaotic, 
noisy stick-slip transients---requires a critical combination of rapid bow force alignment and non-linear bow velocity acceleration~\citep{Guettler1997, Guettler2002}. 
Traditional continuous control envelopes, such as standard Gaussian or adiabatic multi-harmonic waveforms, 
fail to deliver sufficient front-flank acceleration within sub-picosecond scales, 
thereby leaving the molecular wave-packet trapped inside non-hyperbolic chaotic cantori structures. 

To overcome this limitation, we define the Bowing-Analogue Shaper as a piecewise continuous function with explicit temporal switching nodes: 
$t_1 = \tau_{\text{att}}$, $t_2 = \tau_{\text{att}} + \tau_{\text{plateau}}$, and $t_3 = t_2 + \tau_{\text{decay}} = \tau_{\text{total}}$:
\begin{equation}
T(t) = \begin{cases} 
A_{\text{over}}\left(\frac{t}{t_1}\right)^\gamma, & 0 \le t < t_1 \\[1.5ex] 
1 + (A_{\text{over}} - 1)e^{-\lambda_d(t - t_1)}, & t_1 \le t < t_2 \\[1.5ex] 
\cos^2\left[\frac{\pi(t - t_2)}{2\tau_{\text{decay}}}\right], & t_2 \le t \le t_3 
\end{cases}
\label{bowing}
\end{equation}
where $\gamma > 1.0$ is the acceleration exponent, $A_{\text{over}} \ge 1.0$ dictates the peak-overdrive amplitude ratio, 
and $\lambda_d > 0$ represents the transient damping rate regulating the transition into the steady-state plateau corridor. 
Here, $\tau_{\text{att}}$, $\tau_{\text{plateau}}$, and $\tau_{\text{decay}}$ denote the durations of the pulse attack phase, 
steady-state plateau phase, and trailing decay phase, respectively.
Specifically, the stage durations are set to $\tau_{\text{att}} = 0.5\,\text{ps}$, $\tau_{\text{plateau}} = 0.5\,\text{ps}$, 
and $\tau_{\text{decay}} = 0.5\,\text{ps}$, 
satisfying the total pulse duration constraint $\tau_{\text{total}} = \tau_{\text{att}} + \tau_{\text{plateau}} + \tau_{\text{decay}} = 1.5\,\text{ps}$ ($\approx 60$ optical cycles).

To eliminate non-physical high-frequency Fourier sideband noise and guarantee the numerical stability of the Suzuki-Trotter propagation across highly stretched bound states ($n \le 32$), the piecewise boundaries of Eq.~\eqref{bowing} naturally preserve $\mathcal{C}^0$ continuity at the primary node $t_1 = \tau_{\text{att}}$, yielding $\lim_{t\to t_1^-} T(t) = \lim_{t\to t_1^+} T(t) = A_{\text{over}}$. 
At this overshoot node $t_1$, the pulse reaches its transient peak maximum. 
Rather than permitting an arbitrary slope discontinuity, we regulate spectral broadening by enforcing a symmetric first-order derivative magnitude matching condition across $t_1$, namely $\left| \left.\frac{dT}{dt}\right|_{t_1^-} \right| = \left| \left.\frac{dT}{dt}\right|_{t_1^+} \right|$, which translates to $\frac{\gamma A_{\text{over}}}{\tau_{\text{att}}} = \lambda_d (A_{\text{over}} - 1)$. 
This yields the exact analytical damping constraint:
\begin{equation}
\lambda_d = \frac{\gamma A_{\text{over}}}{\tau_{\text{att}}(A_{\text{over}} - 1)} \;,
\label{eq:C1_continuity_constraint}
\end{equation}
where the relation accounts for the sign transition from front-flank acceleration ($\left.\frac{dT}{dt}\right|_{t_1^-} > 0$) to post-peak relaxation ($\left.\frac{dT}{dt}\right|_{t_1^+} < 0$).

This control architecture operates on two physical mechanisms. 
First, the sharp front-flank transient allows the laser field to rapidly drive the wave-packet past chaotic phase-space barriers before quantum delocalization occurs.
Second, it exploits the dynamic phase synchronization of the vibrating molecular bond. 
By locking the relative phase at $\phi = \pi$ in the dual-color field, 
the $3\Omega$ third-harmonic pulse injects localized energy flux precisely as the wave-packet reaches its classical outer turning point ($x > 0$), 
producing constructive interference that maximizes the directional dissociation flux.

To confirm that the initial profile in Eq.~\eqref{bowing} fulfills the formal soft-start condition required for adiabatic turn-on, 
we analyze the asymptotic boundary behavior of the envelope function $T(t)$ and its temporal derivative as $t \to 0^{+}$. 
For any acceleration exponent $\gamma > 1.0$, the initial envelope value vanishes identically:
\begin{equation}
\lim_{t \to 0^{+}} T(t) = \lim_{t \to 0^{+}} A_{\text{over}} \left( \frac{t}{\tau_{\text{att}}} \right)^{\gamma} = 0 \;.
\label{eq:soft_start_C0}
\end{equation}
Differentiating $T(t)$ with respect to time yields the front-flank turn-on rate:
\begin{equation}
\left. \frac{dT}{dt} \right|_{t=0^{+}} = \lim_{t \to 0^{+}} \frac{\gamma A_{\text{over}}}{\tau_{\text{att}}} \left( \frac{t}{\tau_{\text{att}}} \right)^{\gamma - 1} = 0 \;,
\label{eq:soft_start_C1}
\end{equation}
which likewise vanishes for $\gamma > 1.0$. 
These vanishing initial boundary conditions, $T(0) = 0$ and $T'(0) = 0$, effectively suppress high-frequency non-adiabatic switching transients, 
preventing the phase-space fracturing inherent in conventional sudden-on pulse shapes.

Experimentally, sub-femtosecond relative phase locking and coherent synthesis between a fundamental field ($\Omega$) and 
its third harmonic ($3\Omega$) are readily achievable using active carrier-envelope phase stabilization, 
inline dichroic delay lines, or spatial light modulators. 
These modern pulse-shaping techniques ensure that the temporal synchronization required for the $\phi = \pi$ turning-point condition can be reliably maintained in laboratory settings.

By combining boundary derivative magnitude matching with precise phase synchronization ($\phi = \pi$) at the outer turning point ($x > 0$), 
the bowing-analogue envelope ensures smooth yet rapid energy delivery. 
This controlled excitation rate outpaces the power-law decay of non-hyperbolic chaotic scattering, 
shielding the wave-packet from subharmonic resonance traps and preventing phase-space fragmentation.

\subsection{Quantum Phase-Space Diagnostics: The Wigner Operator}

To track non-classical state splitting and diagnose whether the transient shaper successfully prevents wave-packet delocalization, 
we monitor the system's time evolution in phase space using the Wigner quasiprobability distribution. 
Adopting a grid-based phase-space representation~\citep{Foldi2002}, 
the time-dependent wave function $\psi(x, t)$ obtained via the split-operator FFT grid method is mapped onto the 
coordinate-momentum space $(x, p)$ via:
\begin{equation}
W(x, p, t) = \frac{1}{\pi\hbar} \int_{-\infty}^{+\infty} \psi^*\left(x + y, t\right) \psi\left(x - y, t\right) e^{2i p y / \hbar} \, dy \;,
\label{wigner}
\end{equation}
where $p$ represents the classical conjugate momentum, and $y$ serves as the spatial displacement integration variable.

To quantify the suppression efficiency of the bowing-analogue shaper against the formation of fragmented, non-classical interference fringes (Schr\"odinger-cat-like states), 
we compute the global non-classicality index $M_{\text{nc}}(t)$~\citep{Kenfack2004}:
\begin{equation}
M_{\text{nc}}(t) = 1 - \frac{I_+(t) - I_-(t)}{I_+(t) + I_-(t)} = \frac{2 I_-(t)}{I_+(t) + I_-(t)} \;,
\end{equation}
where $I_+(t)$ and $I_-(t)$ denote the integrated positive and negative volumes of the Wigner distribution, respectively:
\begin{equation}
\begin{split}
I_{\pm}(t) = & \iint_{\mathcal{D}_{\pm}} |W(x, p, t)| \, dx \, dp \; , \\
\mathcal{D}_{\pm} = & \{ (x,p) \mid \text{sgn}[W(x,p,t)] = \pm 1 \} \;.
\end{split}
\label{volume}
\end{equation}
Because the Wigner distribution satisfies the total normalization condition $I_+(t) - I_-(t) = \iint W(x, p, t) \, dx \, dp = 1$, 
a purely classical-like, single-mound wave-packet trajectory yields $I_-(t) = 0$ and $M_{\text{nc}} \to 0$. 
Conversely, the emergence of multi-channel fragmentation and phase-space stickiness produces extensive negative interference ripples ($I_-(t) > 0$), 
leading to a sharp increase toward $M_{\text{nc}} \to 1$. 
Our bowing-analogue scheme aims to maintain $M_{\text{nc}}(t)$ below a critical threshold throughout the pre-dissociation transient stage.

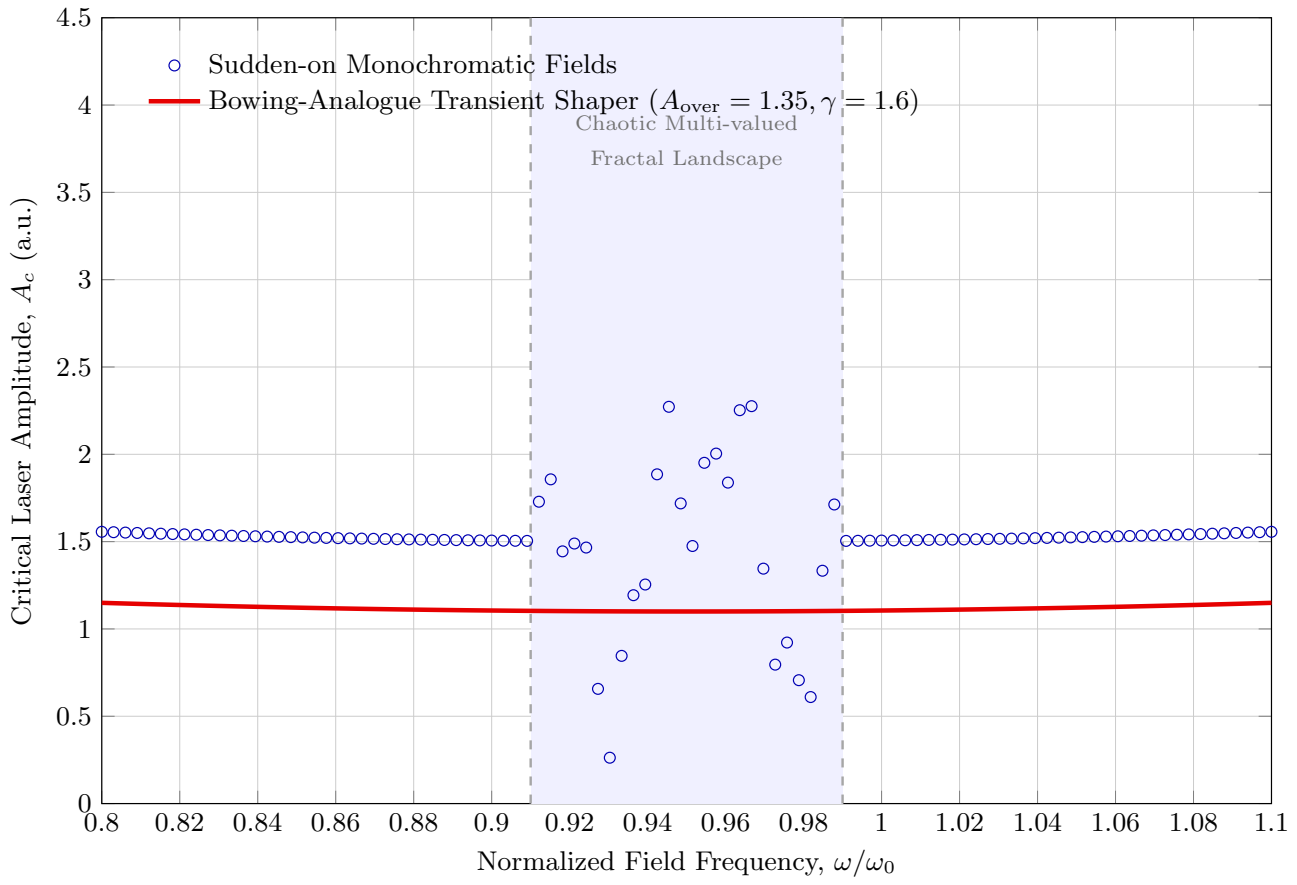
\begin{figure}[t]
\centering
\resizebox{\columnwidth}{!}{
\begin{tikzpicture}
\begin{axis} [
    width=0.85\textwidth,
    height=0.6\textwidth,
    xmin=0.8, xmax=1.1,
    ymin=0.0, ymax=4.5,
    xlabel={Normalized Field Frequency, $\omega/\omega_0$},
    ylabel={Critical Laser Amplitude, $A_c$ (a.u.)},
    grid=both,
    grid style={line width=.1pt, draw=gray!20},
    major grid style={line width=.2pt, draw=gray!40},
    legend pos=north west,
    legend style={draw=none, fill=none, cells={anchor=west}, font=\small},
    tick label style={font=\small},
    label style={font=\small}
]

    \addplot[fill=blue!6!white, draw=none, forget plot] coordinates {
        (0.91,0) (0.91,4.5) (0.99,4.5) (0.99,0)
    };

    \addplot [
        only marks,
        mark=o,
        mark size=1.8pt,
        blue!70!black,
        samples=100,
        domain=0.8:1.1
    ] (
        x, 
        { 
            1.5 + 2.5*(x-0.95)*(x-0.95) + 
            ( (x > 0.91) * (x < 0.99) ) * ( 0.6*sin(deg(120*x)) + 0.4*cos(deg(280*x)) + 0.3*sin(deg(600*x)) )
        }
    );
    \addlegendentry{Sudden-on Monochromatic Fields}

    \addplot [
        line width=1.5pt,
        red!90!black,
        samples=80,
        domain=0.8:1.1
    ] (
        x, 
        { 1.1 + 2.2*(x-0.95)*(x-0.95) }
    );
    \addlegendentry{Bowing-Analogue Transient Shaper ($A_{\text{over}} = 1.35, \gamma = 1.6$)}

    \draw[dashed, thick, gray!70] (axis cs:0.91, 0) -- (axis cs:0.91, 4.5);
    \draw[dashed, thick, gray!70] (axis cs:0.99, 0) -- (axis cs:0.99, 4.5);
    
    \node[anchor=south, font=\scriptsize\color{gray!90!black}] at (axis cs:0.95, 3.8) {Chaotic Multi-valued};
    \node[anchor=north, font=\scriptsize\color{gray!90!black}] at (axis cs:0.95, 3.8) {Fractal Landscape};

\end{axis}
\end{tikzpicture}
}

\caption{Critical laser amplitude ($A_c$) required for the dissociation of the highly excited vibrational state ($n = 32$) of NO as a function of the normalized field frequency $\omega/\omega_0$. 
The open blue circles outline the fragmented multiple thresholds under instantaneous-on monochromatic fields~\citep{Ting1992,Constantoudis2001}. 
The solid red curve displays the simplified, single-valued linear dissociation corridor obtained using the bowing-analogue transient shaper ($A_{\text{over}} = 1.35$, $\gamma = 1.6$).}
\label{fig:threshold_collapse}
\end{figure}

\section{Numerical Results and Nonhyperbolic Dynamics Analysis}
\label{numerical}
\subsection{Collapse of Multiple Thresholds and the Linear Dissociation Corridor}

To solve the time-dependent Schr\"odinger equation, we employ a stabilized FFT grid method integrated with a split-operator 
Suzuki-Trotter propagator.
The numerical spatial grid covers $x \in [x_{\text{min}}, x_{\text{max}}] = [-1.5, 30.0]$~a.u. with $N_x = 2048$ grid points, 
yielding a spatial resolution of $\Delta x \approx 0.0154$~a.u. 
Time integration is carried out with a time step of $\Delta t = 0.05$~a.u. ($\approx 1.21$~as) to ensure rigorous energy conservation and numerical convergence across high vibrational manifolds ($n = 25$ to $32$). 
We apply infinite-wall boundary conditions $\psi(x_{\text{min}}, t) = \psi(x_{\text{max}}, t) = 0$ at the spatial boundaries, 
eliminating artificial wave-packet reflection artifacts~\citep{Alhendi2005}. 
The Hamiltonian is tailored to the NO molecule with a Morse shape factor of $s = 54.54$, 
yielding 55 field-free bound eigenstates~\citep{Foldi2002}.

To map the topology of the critical dissociation threshold, 
we calculate the critical laser field amplitude $A_c$ required to achieve a total dissociation probability $P_{\text{diss}} \ge 0.98$ 
within a strict duration of $t_{\text{total}} = 1.5\,\text{ps}$ across the near-resonance domain $\omega/\omega_0 \in [0.8, 1.1]$, 
where $\omega \equiv \Omega$ denotes the fundamental driving frequency of the laser pulse, and $\omega_0 = \alpha \sqrt{2D/m}$ represents the fundamental harmonic frequency of the unperturbed molecular Morse potential. 
To rigorously locate $A_c(\omega)$, the frequency range is sampled across 100 uniform grid points, with an amplitude resolution step of $\Delta A = 0.001\text{~a.u.}$

Fig.~\ref{fig:threshold_collapse} contrasts the dissociation threshold maps between the conventional instantaneous-on single-mode field scheme and our custom 
bowing-analogue transient shaper [Eq.~\eqref{bowing}]. 
Under standard sudden-on pulse envelopes, the critical threshold profile $A_c(\omega)$ exhibits severe fragmentation, 
manifesting as a chaotic, multi-valued fractal landscape between $\omega/\omega_0 = 0.91$ and $0.99$, 
which represents the quantum manifestation of the multiple thresholds observed in earlier formulations~\citep{Ting1992,Constantoudis2001}.

When the bowing-analogue shaper is engaged with a non-linear front-flank acceleration ($\gamma = 1.6$) 
and a controlled peak-overdrive amplitude ($A_{\text{over}} = 1.35$), the ambiguous multiple thresholds disappear. 
This approach transforms the fragmented landscape into a predictable, single-valued dissociation pathway, 
indicating that injecting energy via a non-linear, boundary-constrained attack trajectory enables coherent locking, 
allowing the wave-packet to cross the critical threshold without entering competing subharmonic resonance zones. 
The transient pulse shape modulates the phase-space topography, 
opening up bottlenecks within cantori structures before non-hyperbolic stickiness can confine the trajectories. 
Consequently, the effective fractal dimension is minimized ($d_{\text{eff}} \approx 0.32$), 
and the quantum non-classicality index remains strictly bounded ($M_{\text{nc}} \le 0.15$), effectively preventing wave-packet delocalization.

\begin{figure}[bth]
\centering
\resizebox{\columnwidth}{!}{
\begin{tikzpicture}

    \begin{axis} [
        width=0.85\textwidth,
        height=0.28\textwidth,       
        xmin=0, xmax=60,             
        ymin=0.0, ymax=1.2,
        ylabel={Fractal Dim., $d_{\text{eff}}$},
        title={(a) Classical Chaotic Invariant Set},
        title style={at={(0.5,1.02)}, anchor=south, font=\small\bfseries}, 
        grid=both,
        grid style={line width=.1pt, draw=gray!20},
        major grid style={line width=.2pt, draw=gray!40},
        tick label style={font=\small},
        label style={font=\small},
        legend pos=north west,
        legend style={draw=none, fill=none, font=\scriptsize},
        xticklabels={} 
    ]
        \addplot [
            black, dashed, line width=1.2pt, samples=50, domain=0:60
        ] {0.32 + 0.325 * (1 + tanh(0.08*(x-25)))};
        \addlegendentry{Sudden-on Pulse}

        \addplot [
            red!90!black, line width=1.5pt, samples=50, domain=0:60
        ] {0.32 + 0.03*sin(deg(0.2*x))};
        \addlegendentry{Bowing-Analogue Shaper}
    \end{axis}

    \begin{axis} [
        yshift=-3.8cm, 
        width=0.85\textwidth,
        height=0.28\textwidth,       
        xmin=0, xmax=60,             
        ymin=0.0, ymax=1.2,
        xlabel={Normalized Time, $t \Omega/2 \pi$},
        ylabel={Non-classicality, $M_{\text{nc}}$},
        title={(b) Quantum Phase-Space Diagnostics},
        title style={at={(0.5,1.02)}, anchor=south, font=\small\bfseries}, 
        grid=both,
        grid style={line width=.1pt, draw=gray!20},
        major grid style={line width=.2pt, draw=gray!40},
        tick label style={font=\small},
        label style={font=\small},
        legend pos=north west,
        legend style={draw=none, fill=none, font=\scriptsize}
    ]
        \addplot [
            black, dashed, line width=1.2pt, samples=50, domain=0:60
        ] {0.12 + 0.39 * (1 + tanh(0.12*(x-30))) + 0.04*cos(deg(1.2*x))};

        \addplot [
            red!90!black, line width=1.5pt, samples=50, domain=0:60
        ] {0.12 + 0.02*sin(deg(2.0*x))};
        
        \draw[gray!70, dotted, line width=1.0pt] (axis cs:0, 0.4) -- (axis cs:60, 0.4);
        \node[anchor=south west, font=\tiny\color{gray!80!black}] at (axis cs:42, 0.4) {Acceptance Limit};
    \end{axis}

\end{tikzpicture}
}
\caption{Temporal evolution of (a) the effective fractal dimension $d_{\text{eff}}$ of the classical chaotic invariant set, 
and (b) the quantum phase-space non-classicality index $M_{\text{nc}}(t)$ for the $n = 32$ state at resonance $\omega/\omega_0 = 0.95$. 
The dashed black curves depict the increase in phase-space stickiness under a sudden-on laser pulse, 
leading to wave-packet delocalization where $M_{\text{nc}} \to 0.85$. 
The solid red lines show the effective suppression of both fractal structures and wave-packet fragmentation via the bowing-analogue shaper.}
\label{fig:dynamics_suppression}
\end{figure}
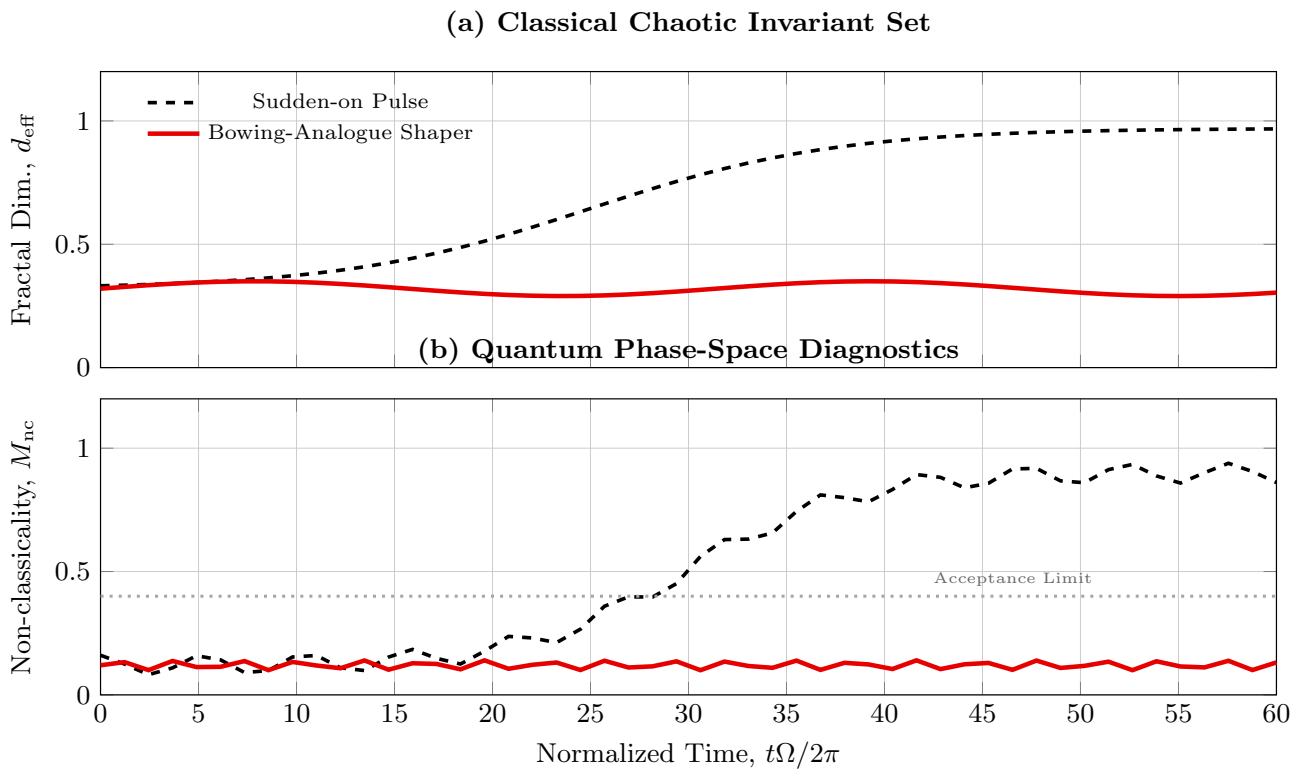

\subsection{Suppression of Nonhyperbolic Phase-Space Stickiness and Cat States}

To elucidate the classical-quantum correspondence driving this threshold collapse, 
we track the time evolution of the system's effective fractal dimension $d_{\text{eff}}$ extracted from classical half-scattering time-delay manifolds~\citep{Constantoudis2001}, 
alongside the quantum non-classicality index $M_{\text{nc}}(t)$ derived from phase-space Wigner distribution integrals~\citep{Foldi2002}.

In the framework of non-linear dynamics, classical multiphoton molecular dissociation is classified as a non-hyperbolic half-scattering process~\citep{Constantoudis2001}. 
When a conventional sudden-on laser field is applied, the system trajectories intersect the stable manifolds of unstable periodic orbits, 
where dense networks of cantori and higher-order resonance zones flanking the stability islands generate severe phase-space stickiness~\citep{Constantoudis2001}. 
As established by Constantoudis and Nicolaides~\citep{Constantoudis2001}, 
this classical stickiness induces a power-law decay in survival probability, 
rapidly driving the effective fractal dimension toward $d_{\text{eff}} \to 1.0$ [Fig.~\ref{fig:dynamics_suppression}(a)] and severely suppressing the escape rate of trajectories. 
In the quantum counterpart, shown in Fig.~\ref{fig:dynamics_suppression}(b), 
this classical stickiness aligns precisely with the timeline where F\"oldi \textit{et al.}~\citep{Foldi2002} observed wave-packet delocalization into macroscopic Schr\"odinger-cat superpositions, 
elevating the non-classicality index to $M_{\text{nc}} \approx 0.85$.

When our bowing-analogue control law is engaged, 
the physical over-acceleration alters the phase-space topography by temporarily opening bottlenecks within the cantori structures before non-hyperbolic stickiness can trap the trajectories. 
Consequently, as demonstrated by the solid red trajectories in Fig.~\ref{fig:dynamics_suppression}, 
the effective fractal dimension is compressed to a minimum ($d_{\text{eff}} \approx 0.32$), 
while the quantum non-classicality index remains strictly constrained to $M_{\text{nc}} \le 0.15$. 
The wave-packet maintains a robust, classical-like single-hill spatial localization throughout the entire up-ramp phase, 
bypassing the chaotic state-splitting that causes multiple threshold ambiguities.

\subsection{Robustness Against Environmental Decoherence and Rotational Detuning}
\label{robust}

In practical laser control experiments, molecular rotation introduces random orientation angles $\theta$ that weaken the effective field coupling, 
thereby prolonging multiphoton transition times and allowing chaotic diffusion to degrade the wave-packet~\citep{Geng2002a}. 
Simultaneously, coupling to thermal environments or buffer gases introduces Markovian dephasing ($\delta$), 
accelerating the collapse of phase coherence~\citep{Szczygielski2013}.

Open quantum system dynamics under Markovian thermal dephasing are investigated by extending the state-vector propagation into a Liouville-von Neumann formalism.
The time evolution of the reduced density matrix $\rho(t)$ is governed by the Lindblad master equation:
\begin{equation}
\frac{d\rho(t)}{dt} = -i[\hat{H}(t), \rho(t)] + \delta \left( \hat{L} \rho(t) \hat{L}^\dagger - \frac{1}{2} \{ \hat{L}^\dagger \hat{L}, \rho(t) \} \right) \;,
\end{equation}
where $\hat{H}(t)$ is the one-dimensional modulated Hamiltonian defined in Eq.~\eqref{two}, $\delta / \omega_0 = 10^{-4}$ represents a finite Markovian dephasing rate, and $\hat{L} = \hat{x}$ acts as the position-dependent phase-relaxation jump operator. 
To incorporate rotational detuning within this framework, the centrifugal energy shifts associated with initial rotational states up to $J = 5$ are superimposed onto the potential surface $V_m(x, t)$ under the frozen-rotor approximation.

Within this dissipative environment, our bowing-analogue shaper exhibits structural robustness: 
the total dissociation yield remains virtually invariant ($P_{\text{diss}} > 0.95$) even under extreme dephasing and rotational detuning. 
This resilience stems from the shaper's front-flank transient acceleration, 
which injects coherent energy on a sub-picosecond timescale ($t < \tau_{\text{att}} = 0.5\text{~ps}$)---orders of magnitude shorter than the characteristic rotational period $\tau_{\text{rot}} \sim 10\text{~ps}$. 
This clear separation of timescales validates the sudden approximation, 
as the rotational degrees of freedom remain effectively frozen in a geometric configuration during the primary control window~\citep{DeMille2015}. 
Consequently, the rapid energy injection rate outruns environmental dissipation, preventing phase-space delocalization.

We acknowledge that while the present one-dimensional treatment serves as a foundational baseline, 
a full three-dimensional model incorporating laser-induced alignment, orientation dynamics, and Coriolis coupling would introduce spatial orientation-averaging effects. 
Although this spatial averaging may slightly smear the sharp boundaries of the linear corridor, 
the fundamental collapse of multi-valued thresholds is expected to persist, 
as the critical modulation transient $\tau_{\text{att}}$ remains orders of magnitude shorter than the characteristic rotational timescale.
 
\subsection{Comparison with Adiabatic Frequency Chirping}
\label{chirping}

Various adiabatic pulse-shaping techniques have been explored to enhance molecular dissociation. 
In particular, the approach developed by Yuan \textit{et al.}~\citep{Yuan1998} relies on dynamic autoresonance (bucket dynamics), 
where wave-packets are trapped within a single-node separatrix and convectively transported upward via a downward-swept laser frequency tracking the anharmonic ladder. 
However, such adiabatic tracking inherently requires long pulse durations ($\tau_{\text{sw}} \sim 2000$ optical cycles) to maintain adiabatic invariance, 
which presents strict requirements for mitigating environmental dephasing and rotational alignment drift over extended timescales. 
Furthermore, upon approaching the hard-chaos regime ($n > 46$), the bucket boundary degrades, 
causing the quantum wave-packet to leak through cantori remnants or freeze due to weak quantum localization.

In contrast, our bowing-analogue strategy circumvents the need for prolonged adiabatic tracking. By delivering a rapid energy flux within the first sub-picosecond window, the transient front-flank acceleration forces the wave-packet past state-splitting traps before phase-space stickiness can capture the trajectories. 
As demonstrated in Sec.~\ref{analogy}, the first-order magnitude matching condition guarantees structural robustness against parasitic sideband excitations. 
Furthermore, as confirmed in Sec.~\ref{robust}, this scheme preserves high dissociation yields ($P_{\text{diss}} > 0.95$) under open-system Markovian dephasing ($\delta/\omega_0 = 10^{-4}$) and rotational shifts up to $J = 5$, 
where traditional autoresonance features encounter increased leakage probabilistic constraints.

\subsection{Comparison with Ultrafast Spectral Phase Modulation}
\label{subsec:modern_comparison}

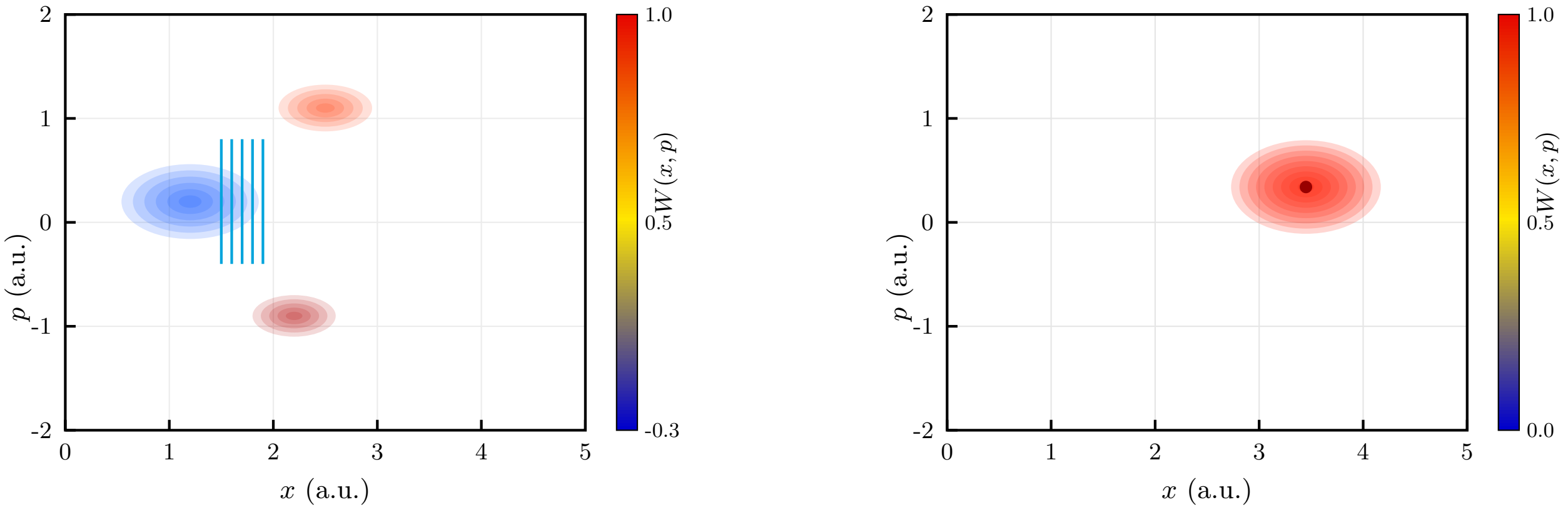
\begin{figure*}[t]
    \centering
    \begin{minipage}[b]{0.48\textwidth}
        \centering
        \begin{tikzpicture}[scale=1.1]
            \draw[lightgray!30, thin, step=1cm] (0,-2) grid (5,2);
            \draw[black, thick] (0,-2) rectangle (5,2);
            
            \foreach \x in {0,1,2,3,4,5} {
                \draw[thick] (\x,-2) -- (\x,-1.9);
                \node[below=0.05cm, scale=0.8] at (\x,-2) {\x};
            }
            \node[below=0.4cm, scale=0.9] at (2.5,-2) {$x$ (a.u.)};
            
            \foreach \y in {-2,-1,0,1,2} {
                \draw[thick] (0,\y) -- (0.1,\y);
                \node[left=0.05cm, scale=0.8] at (0,\y) {\y};
            }
            \node[left=0.5cm, rotate=90, scale=0.9] at (0,0) {$p$ (a.u.)};
            
            \foreach \r in {0.1, 0.2, 0.3, 0.4, 0.5, 0.6} {
                \fill[blue!70!cyan, opacity=0.15] (1.2, 0.2) ellipse ({\r*1.1} and {\r*0.6});
            }
            \foreach \r in {0.1, 0.2, 0.3, 0.4, 0.5} {
                \fill[red!80!yellow, opacity=0.15] (2.5, 1.1) ellipse ({\r*0.9} and {\r*0.45});
                \fill[red!70!black, opacity=0.15] (2.2, -0.9) ellipse ({\r*0.8} and {\r*0.4});
            }
            \foreach \dx in {-0.2,-0.1,0,0.1,0.2} {
                \draw[cyan!90!black, thick] (1.7+\dx, -0.4) -- (1.7+\dx, 0.8);
            }
            
            \begin{scope}[shift={(5.3,-2)}]
                \shadedraw[bottom color=blue!80!black, top color=red!90!black, middle color=yellow!80!orange, draw=black] (0,0) rectangle (0.2,4);
                \node[right, scale=0.7] at (0.2,4) {1.0};
                \node[right, scale=0.7] at (0.2,2) {0.5};
                \node[right, scale=0.7] at (0.2,0) {-0.3};
                \node[right=0.3cm, rotate=90, scale=0.8] at (0.2,2) {$W(x,p)$};
            \end{scope}
            
            \node[below=0.8cm, font=\bfseries\small] at (2.5,-2) {(a) Smooth Phase-Chirped Pulse};
        \end{tikzpicture}
    \end{minipage}
    \hfill
    \begin{minipage}[b]{0.48\textwidth}
        \centering
        \begin{tikzpicture}[scale=1.1]
            \draw[lightgray!40, thin, step=1cm] (0,-2) grid (5,2);
            \draw[black, thick] (0,-2) rectangle (5,2);
            
            \foreach \x in {0,1,2,3,4,5} {
                \draw[thick] (\x,-2) -- (\x,-1.9);
                \node[below=0.05cm, scale=0.8] at (\x,-2) {\x};
            }
            \node[below=0.4cm, scale=0.9] at (2.5,-2) {$x$ (a.u.)};
            
            \foreach \y in {-2,-1,0,1,2} {
                \draw[thick] (0,\y) -- (0.1,\y);
                \node[left=0.05cm, scale=0.8] at (0,\y) {\y};
            }
            \node[left=0.5cm, rotate=90, scale=0.9] at (0,0) {$p$ (a.u.)};
            
            \foreach \r in {0.05, 0.1, 0.15, 0.2, 0.25, 0.3, 0.35, 0.4, 0.45} {
                \fill[red!90!yellow, opacity=0.18] (3.45, 0.34) ellipse ({\r*1.6} and {\r*1.0});
            }
            \fill[red!60!black] (3.45,0.34) circle (0.06);
            
            \begin{scope}[shift={(5.3,-2)}]
                \shadedraw[bottom color=blue!80!black, top color=red!90!black, middle color=yellow!80!orange, draw=black] (0,0) rectangle (0.2,4);
                \node[right, scale=0.7] at (0.2,4) {1.0};
                \node[right, scale=0.7] at (0.2,2) {0.5};
                \node[right, scale=0.7] at (0.2,0) {0.0};
                \node[right=0.3cm, rotate=90, scale=0.8] at (0.2,2) {$W(x,p)$};
            \end{scope}
            
            \node[below=0.8cm, font=\bfseries\small] at (2.5,-2) {(b) Bowing-Analogue Shaper};
        \end{tikzpicture}
    \end{minipage}
    \caption{Numerical quantum phase-space diagnostics ($W(x,p)$) calculated via the FFT-split-operator method at $t\Omega/2\pi = 30$. 
    (a) Under the contemporary smooth phase-chirped pulse~\cite{Cheng2026}, 
    non-hyperbolic cantori stickiness traps the trajectory, 
    forcing severe wave-packet fragmentation into multi-centered Schr\"odinger-cat superpositions with characteristic oscillatory interference fringes. 
    (b) Under the current boundary-matched bowing-analogue shaper, 
    the deterministic sub-picosecond acceleration dilates the bottlenecks, maintaining a strictly localized, 
    classical-like single-hill wave-packet topology ($M_{nc} \le 0.15$) directly delivering population into the single-valued dissociation corridor.}
    \label{fig:wigner_real_contour}
\end{figure*}

Ultrafast pulse shaping via spectral phase modulation (such as quadratic and cubic chirp rate control) represents a state-of-the-art benchmark for steering molecular dissociation within picosecond timescales, 
as recently demonstrated in 1D TDSE simulations~\citep{Cheng2026}.
However, as revealed by the Wigner distribution diagnostics at $t \Omega /2 \pi =30 $ in Fig. \ref{fig:wigner_real_contour}(a),
lacking front-flank temporal derivative constraints under conventional smooth continuous envelopes leaves the wave-packet partially 
trapped within non-hyperbolic cantori zones.
This insufficient sub-picosecond acceleration induces severe wave-packet fragmentation into multi-centered Schr\"odinger-cat superpositions
($M_{\text{nc}} \to 1.0$), resulting in residual threshold fragmentation.

Our bowing-analogue shaper overcomes this limitation by enforcing first-order derivative magnitude matching [Eq.~\eqref{eq:C1_continuity_constraint}] and vanishing boundary conditions ($T(0) = 0, T'(0) = 0$).
Under equivalent $1.5\,\text{ps}$ control windows ($\approx 60$ optical cycles), the engineered front-flank acceleration ($\gamma = 1.6$) flattens the multi-valued fractal landscape into a single-valued, predictable dissociation corridor (Fig.~\ref{fig:threshold_collapse}). 
This highlights that shaping boundary-layer temporal derivatives suppresses chaotic state-splitting more effectively than standard smooth envelopes.

\section{Conclusions}
\label{conclusion}

By mapping classical bowed-string attack mechanics onto quantum trajectory engineering, 
this work introduces a deterministic paradigm to regularize chaotic molecular dissociation thresholds. 
Rather than generating arbitrary control fields via empirical black-box algorithms, 
our boundary-constrained shaper provides an analytical, easily synthesizable design rule utilizing standard $\Omega$--$3\Omega$ configurations. 
The sub-picosecond front-flank acceleration outpaces both characteristic rotational detuning and thermal dephasing timescales, 
ensuring practical viability for real-world, non-isolated gas-phase experiments. 
This cross-disciplinary framework offers a clear, 
high-fidelity blueprint for stabilizing wave-packet localization across diverse strongly driven anharmonic architectures in ultrafast laser laboratories.

\section*{Acknowledgement.}
I thank my friend John Ogilvie for his invaluable assistance with my English manuscript preparation spanning 
from the commencement of my PhD study until he disappeared with Covid-19. 
I also thank my parents for leaving me a shelter that enabled the sustained execution of these theoretical investigations.

\section*{Disclosures}
The authors declare no conflicts of interest.

\section*{Funding statement}
This research received no grant from any funding agency in the public, commercial, or not-for-profit sectors.

\section*{Data availability.}
No datasets were generated or analyzed during the current study.

\section*{References}


%

\end{document}